\documentclass[12pt]{iopart}

\usepackage{bm}% bold math
\def\trade{{\bigcirc}\!\!\!\!\!\mbox{{\tiny R}}}
\def\mathmath{{\it Mathematica}$^{\trade}$\,\,}
\advance\voffset by +25pt
\begin{document}

\title[Stark effect for linear potentials]{The Stark effect in linear potentials}

\author{R W Robinett$^1$} 

\address{$^1$ Department of Physics,
The Pennsylvania State University,
University Park, PA 16802 USA}
\ead{rick@phys.psu.edu}

\begin{abstract}
We examine the Stark effect (the second-order shifts in the
energy spectrum due to an external constant force) for two 1-dimensional
model quantum mechanical 
systems described by linear potentials, the so-called quantum bouncer
(defined by $V(z) = Fz$ for $z>0$ and $V(z) = \infty$ for $z<0$) 
and the symmetric linear potential (given by $V(z) = F|z|$). We show
how straightforward use of the most obvious properties of the Airy function 
solutions and simple Taylor expansions give closed form results 
for the Stark shifts in both systems. These exact results
are then compared to other approximation techniques, such as perturbation 
theory and  WKB methods. These expressions add to the small number of 
closed-form descriptions available for the Stark effect in model quantum 
mechanical systems.
\end{abstract}


% Explanations for the PACS numbers
% 02.30.Gp -- Special Functions
% 03.65.Ge -- Solutions of wave equations: bound states
% 03.65.Sq -- Semiclassical theories and applications

\maketitle

\section{Introduction}

There are only a handful of model quantum mechanical systems which
admit exact solutions, and from which information is easily extracted in
closed-form expressions. The most familiar examples include 
the Kepler problem (Coulomb potential 
for the hydrogen atom), the harmonic oscillator 
(for vibrational states), spherical harmonics (for rotational motion), 
and the infinite square well. These examples all allow for a relatively
complete analysis of the mathematical descriptions of the position-space
wavefunctions, including closed-form expressions for 
normalizations, expectation values, and even
matrix elements for use in applications such as sum rules and perturbation
theory.

The so-called `quantum bouncer' system, defined by the potential
\begin{equation}
V(z) = \left\{
\begin{array}{cc}
Fz     & \mbox{for $0 \leq z$} \\
\infty & \mbox{for $z<0$}
\end{array}
\right.
\, ,
\label{quantum_bouncer_potential}
\end{equation}
has been a staple of the pedagogical literature, in both
articles \cite{bouncer_1}  - \cite{bouncer_5}
and textbooks
\cite{bouncer_6} - \cite{bouncer_9}. 
It not only has solutions
given by well-studied special functions 
(Airy functions \cite{airy_function_monograph}), but it can also 
be analyzed using a variety of approximation methods for
comparison to exact results. Most importantly, however, 
this 'academic' problem has received renewed attention 
as the simplest model for the {\it quantum states of neutrons in the Earth's
gravitational field} \cite{neutron_bound_states}. 
The linear potential plus infinite wall has also been used to analyze 
surface-Landau-level resonance data \cite{wanner} where electrons
are trapped between the surface potential  and a weak magnetic force
given by $F=e v_F H/c$. In addition, 
related recent experimental realizations of motion in such a `bouncer'
potential also 
include systems of atoms \cite{cesium}, Bose-Einstein condensates
\cite{bec}, and even optical analogs (`photon bouncing balls') 
\cite{bouncing_photon}.
These proverbial 'real life' applications help to promote the quantum 
bouncer to the small set of important physical realizations of 
mathematically soluble problems which are accessible to students 
of quantum mechanics at the undergraduate level.

A simple extension of this problem, the symmetric linear potential,
given by $V(z) = F|z|$, also shares Airy function solutions. Moreover,
it has the same symmetry property as systems 
such as the harmonic oscillator or symmetric infinite well and so
can be discussed in a similar context, including using concepts such as
parity. For example, 
all three cases can be thought of as belonging to the class 
of power-law potentials of the form
\begin{equation}
V^{(k)}(z) = V_0 \left|\frac{z}{a}\right|^{k}
\label{power_law}
\end{equation}
with $k = 1,2$ and $\infty$ for the three systems respectively. 
One can then systematically explore the behavior of energy eigenvalues
\cite{sukhatme} or even higher order effects, such as the Stark shifts
\cite{polar}, and how they depend on the power-law exponent ($k$),
quantum number, and physical parameters, such as the mass.

The Stark effect, the response of the quantized energy levels to the 
addition of a constant  external force, is another staple of quantum
mechanics texts, with closed form results possible in a number of the
same tractable systems mentioned above. For example, the addition of 
a linear potential of the form $\overline{V}(z) = \overline{F}z$ to the
harmonic oscillator (HO) can be easily accommodated with an 
exact solution (requiring only a simple '{\it complete the squares}' trick), 
leading to a second order shift,
\begin{equation}
E_n^{(2)}(HO) = - \frac{\overline{F}^2}{2m\omega^2}
\, , 
\label{harmonic_oscillator_stark}
\end{equation}
namely a constant, lowering the energies of all of the states. 
For the symmetric infinite well (SIW), with walls at $x=\pm a$,
methods involving inhomogeneous differential equations 
\cite{mavromatis_1} (the so-called Dalgarno-Lewis
method \cite{dalgarno}) and perturbation theory summation approaches  
\cite{mavromatis_2}, \cite{mavromatis_3}
have been used to evaluate the second order shift, yielding
\begin{equation}
E_{n}^{(2)}(SIW) = 
\frac{\overline{F}^2a^2}{12 E_n^{(0)}}
\left[ 1- \frac{15}{(n+1)^2\pi^2}\right]
\, , 
\label{infinite_well_stark}
\end{equation}
where
\begin{equation}
E_n^{(0)} \equiv \frac{\hbar^2 \pi^2(n+1)^2}{8ma^2}
\qquad
\mbox{where \, $n = 0,1,2,...$}
\end{equation}
define the unperturbed energies in an infinite well of width $2a$. 
In contrast to the harmonic oscillator, the Stark shift here is negative only
for the ground state (as required by second-order perturbation theory) but
positive for all excited states, 
which is an interesting qualitative difference.
It is in this context that we will examine
the Stark shifts for the quantum bouncer and the related symmetric
linear potential. 
Given the simple nature of the Airy function solutions (or rather the
differential equation they satisfy), it's perhaps not 
surprising that exact expressions for the Stark shifts in both systems
can also be derived, and that is the topic of this paper.

In the next section, we review the solutions of the 
quantum bouncer problem, focusing
on providing a list of exact results needed for quantities such as
normalizations, expectation values, and other matrix elements.
We find that the Stark shifts are actually 
trivially obtained in this case. In Sec.~\ref{sec:symmetric_linear_potential},
we provide the corresponding results for the more interesting case of
the symmetric linear potential. For this example, 
we make use of the fact that the simple form of the Airy differential
equation provides dramatic simplifications for the Taylor series expansion
of the exact energy eigenvalue condition for the 'perturbed' case, and
we find closed form expressions for the Stark shifts in this system as well, 
one of our main results. In both
cases, we compare the exact results to WKB approximations for the Stark
shifts, which have been discussed earlier in the literature \cite{polar},
as well as with second-order perturbation theory. 
Finally we put our results into context in Sec.~\ref{sec:conclusions}
where we discuss our conclusions.

\section{Stark effect for the quantum bouncer}
\label{sec:quantum bouncer}

We begin by reviewing the solutions for the quantum bouncer problem,
defined by the potential in Eqn.~(\ref{quantum_bouncer_potential}). 
The Schr\"{o}dinger equation in the region $z>0$ reduces to
\begin{equation}
-\frac{\hbar^2}{2m} \frac{d^2 \psi_n(z)}{dz^2}
+ Fz \psi_n(z) = E_n \psi_n(z)
\, , 
\label{bouncer_se}
\end{equation}
which can be written in the form
\begin{equation}
\psi''_{n}(x) = (x-\beta_n) \psi_{n}(x)
\label{airy_version}
\end{equation}
using the change of variable $z = \rho x$ and the definitions
\begin{equation}
\rho = \left(\frac{\hbar^2}{2mF}\right)^{1/3}
\qquad
\quad
\mbox{and}
\quad
\qquad
\beta_n = \frac{E_n}{F\rho} \equiv \frac{E_n}{{\cal E}_0}
\, . 
\label{energy_eigenvalues}
\end{equation}
The solutions of Eqn.~(\ref{airy_version}) are the two linearly-independent
Airy functions, $Ai(x-\beta_n)$ and $Bi(x-\beta_n)$. (The now
standard reference on the application of Airy functions 
in physics is Ref.~\cite{airy_function_monograph}.) The $Bi$ solution
(which diverges for large positive argument) does not satisfy the boundary
condition $\psi_n(z\rightarrow \infty) = 0$ and so is excluded. 
 The energy eigenvalues are then determined by the boundary
condition imposed by the infinite wall at the origin, namely that
$\psi(z=0) = Ai(-\beta_n)$. The quantized energies are then given in terms
of the zeros of the well-behaved Airy function, $Ai(-\zeta_n)$, with 
$E_n = +\zeta_n {\cal E}_0$. 
In many other systems, the explicit manner in which 
the application of the boundary conditions leads to discrete energies 
is less than mathematically obvious. In this case, however,
one simply `slides' the single $Ai(x)$ solution along the
axis (via the $\beta_n$ shift) 
until one of its zeros coincides with the origin and therefore
satisfies the boundary condition at the infinite wall.
(We note that Eqn.~(\ref{bouncer_se}) can be easily solved in momentum
space, and then connected to the integral representation of the
Airy function \cite{bouncer_9}. The energy eigenvalue conditions for this
problem, and especially our systematic expansion of them, are much more 
easily derived directly from the position space form of the Schr\"odinger
equation.)

Standard handbook results for the asymptotic
behavior of the $Ai(x)$ zeros \cite{handbook} give
\begin{equation}
E_n 
\sim  
{\cal E}_0 \left[ \frac{3\pi}{2}(n-1/4)\right]^{2/3}
= 
{\cal E}_0 \left[ \frac{3\pi}{4}(2n-1/2)\right]^{2/3}
\label{quantum_bouncer_wkb}
\end{equation}
for large $n$ (where the labeling starts with $n=1$.) 
We note that the WKB prediction for the energy eigenvalues
for this system is
\begin{equation}
\!\!\!\!\!\!\!\!\!\!\!\!\!
\int_{0}^{E_{\tilde{n}}/F} \sqrt{2m(E_{\tilde{n}}-Fz)}\,dz = (\tilde{n}+C_L +C_R) \hbar \pi
\qquad
\mbox{with \, $\tilde{n}=0,1,2,...$}
\label{wkb_bouncer}
\end{equation}
and the appropriate values of the matching coefficients are $C_L = 1/2$
(at the left turning point, where there is an infinite wall boundary 
condition) and $C_R = 1/4$ (for the smoother potential at the right turning
point).  Solving for the quantized energies gives
\begin{equation}
E_{\tilde{n}}(WKB) 
= \left[\frac{3\pi (\tilde{n}+3/4)}{2}\right]^{2/3} {\cal E}_0\end{equation}
where $\tilde{n}=0,1,2,...$, which agrees with the `handbook' result in
Eqn.~(\ref{quantum_bouncer_wkb}) for large $\tilde{n}$.

The wavefunctions for positive $z$ are then given by
\begin{equation}
\psi_n(z) = N_n Ai\left(\frac{z}{\rho} - \zeta_n\right)
\label{quantum_bouncer_solutions}
\end{equation}
and the orthonormality properties of these solutions can be investigated very
straightforwardly. For example, 
the normalization constant can be determined in closed form by using
one of the simple identities, Eqn.~(\ref{integral_diagonal_0}),
involving integrals Airy functions collected in the Appendix, 
where we find that 
\begin{equation}
N_n = \frac{1}{\sqrt{\rho} \, Ai'(-\zeta_n)}
\label{normalization}
\,.
\end{equation}
(Gea-Banacloche \cite{bouncing_ball} was the first to find this relation
numerically, and soon afterwards Vall\'{e}e \cite{vallee}
and Goodmanson \cite{goodmanson} demonstrated it analytically. In this work,
we choose $Ai'(-\zeta_n)$ in the normalization instead of $|Ai'(-\zeta_n)|$,
as done in Refs.~\cite{bouncing_ball} - \cite{goodmanson}, in order
to simplify some of the expressions below; 
such a choice of phase has, of course, no physical significance.)
One can then just as easily confirm that $\langle \psi_n | \psi_m \rangle = 0$ 
for $n\neq m$ by using Eqn.~(\ref{integral_off_diagonal_0}).

As noted by Goodmanson \cite{goodmanson}, 
the expectation values of the potential and kinetic energies can be evaluated
 by using the identities in Eqns.~(\ref{integral_diagonal_1}) and
(\ref{integral_momentum}) respectively to find
\begin{eqnarray}
\langle n |V(z)| n \rangle & = &
F \int_{0}^{\infty} z\, |\psi_n(z)|^2\, dz \nonumber \\
& = & \frac{F \rho^2}{\rho[Ai'(-\zeta_n)]^2 } 
\int_{0}^{\infty} x \, Ai(x - \zeta_n)^2\,dx \nonumber \\
& = & (F\rho) \left(\frac{2 \zeta_n}{3}\right) = \frac{2}{3} E_n
\label{potential_energy}
\end{eqnarray}

and
\begin{eqnarray}
\frac{1}{2m} \langle n | \hat{p}^2 |n \rangle
& = & \frac{\hbar^2}{2m}
\int_{0}^{\infty} \left|\frac{d\psi_n(z)}{dz}\right|^2\,dz \nonumber \\
& = & \left( \frac{\hbar^2}{2m\rho^2} \right)
\left[
\frac{\rho}{\rho \, [Ai'(-\zeta_n)]^2}\right]
\int_{0}^{\infty} [Ai'(x-\zeta_n)]^2\,dx \nonumber \\
& = & 
\left(\frac{\hbar^2}{2m\rho^2}\right)
\left(\frac{\zeta_n}{3}\right) = 
\frac{{\cal E}_0 \zeta_n}{3} = \frac{1}{3}E_n
\label{kinetic_energy}
\end{eqnarray}
where an integration by parts has been used in Eqn.~(\ref{kinetic_energy})
to bring it to the form of the integral in Eqn.~(\ref{integral_momentum}).
(One can also make direct use of the Airy differential equation, 
$A'' = zA$, to rewrite $\langle n|\hat{p}^2|n\rangle$ in terms 
of the same integral in Eqn.~(\ref{potential_energy}).)
These calculations confirm that both quantities are consistent 
with the virial theorem.
For future reference, we find that  the dipole matrix elements are given by
\begin{equation}
\langle \psi_n |z| \psi_k\rangle = -\frac{2\rho}{(\zeta_n - \zeta_k)^2}
\label{bouncer_dipole_matrix_elements}
\end{equation}
where we use Eqn.~(\ref{integral_off_diagonal_1}).

Turning now to the Stark effect, we note that the result of a
perturbing potential of the form $\overline{V}(z) = \overline{F}z$
is trivially realized since the form of the solutions remain the same 
with the global substitution
$\tilde{F} \rightarrow (F+\overline{F})$. 
The new energy eigenvalues are thus obtained  
from Eqn.~(\ref{energy_eigenvalues}) as
\begin{equation}
\tilde{E}_n = ({\cal E}_0\zeta_n)\, \left[1+\frac{\overline{F}}{F}\right]^{2/3}
\label{simple_form}
\, . 
\end{equation}
With this simple form, the predictions for the first-, second-, and
third-order energy shifts are given by a simple expansion, namely
\begin{equation}
\!\!\!\!\!\!\!\!\!\!\!\!\!\!\!\!\!\!\!\!
\!\!\!\!\!\!\!\!\!\!\!\!
E_n^{(1)} = \frac{2}{3} \left(\frac{\overline{F}}{F}\right) E_n^{(0)}
\, , 
\qquad
E_n^{(2)} = -\frac{1}{9} \left(\frac{\overline{F}}{F}\right)^2 E_n^{(0)}
\, , 
\quad
\mbox{and}
\quad
E_n^{(3)} = \frac{4}{81} \left(\frac{\overline{F}}{F}\right)^3 E_n^{(0)}
\, .
\end{equation}
The first-order perturbation theory result,
\begin{equation}
E_n^{(1)} = \langle \psi_n |\overline{V}(z)| \psi_n\rangle
= \overline{F} \langle \psi_n |z| \psi_n\rangle
=  \frac{2}{3} \left(\frac{\overline{F}}{F}\right) ({\cal E}_0 \zeta_n)
\, , 
\label{first_order_shift}
\end{equation}
is easily confirmed by the same integral as in Eqn.~(\ref{potential_energy}). 
The second-order energy term is given by the standard expression
\begin{equation}
E_n^{(2)} = \sum_{k \neq n} 
\frac{|\langle n |\overline{V}(z)|k\rangle|^2}{(E_n^{(0)} - E_k^{(0)})}
\label{general_second_order_shift}
\end{equation}
so that the predicted Stark shift for the quantum bouncer is
\begin{equation}
E_n^{(2)} = -4 \left(\frac{\overline{F}}{F}\right)^2
{\cal E}_0 \left[
\sum_{k \neq n} \frac{1}{(\zeta_k - \zeta_n)^5}
\right]
\, . 
\end{equation}
It has been shown in Ref.~\cite{belloni_robinett_airy_sum_rules} that 
this result (and the corresponding double summation 
giving the third-order expression, $E_n^{(3)}$)
provides new constraints on  combinations of zeros of Airy functions, 
which can be easily confirmed numerically using tools such as
\mathmath.   This is one example where novel applications
of standard quantum mechanical methods can provide new results in
 mathematical physics.

We also note that the WKB approximation in Eqn.~(\ref{wkb_bouncer}) has the
same global redefinition of $F \rightarrow \tilde{F}= F + \overline{F}$, 
yielding the same additional factor of $(1+\overline{F}/F)^{2/3}$ 
found in the exact result in Eqn.~(\ref{simple_form}), so that the WKB
approximation gives the correct large $n$ behavior of the first- and
second-order shifts as well.

\section{The symmetric linear potential}
\label{sec:symmetric_linear_potential}

The case of the symmetric linear potential, 
a seemingly small variation on the quantum bouncer potential, namely
\begin{equation}
V(z) = F|z|
\, , 
\end{equation} 
adds a number of interesting features, most notably making the evaluation
of the $\overline{F}$-dependent shifts in energy levels more challenging.
But use of the simplest properties of the Airy function solutions leads
to a straightforward calculational algorithm which can still give closed-form 
results for the energy shifts to any desired order in $\overline{F}/F$. 
We begin by reviewing the solutions of the symmetric linear potential.

The energy eigenstates in this symmetric potential can be classified by
parity, and the odd-parity states are automatically related to those of
the `half-well' problem (namely the quantum bouncer above) by
\begin{equation}
\psi_n^{(-)}(z) =
\left\{
\begin{array}{cc}
+\psi_n(z)/\sqrt{2} & \qquad  \mbox{for $z \geq 0$} \\
-\psi_n(-z)/\sqrt{2} & \qquad \mbox{for $z \leq0$} 
\end{array}
\right.
\end{equation}
where the $\psi_n(z)$ are given by
Eqns.~(\ref{quantum_bouncer_solutions}) and (\ref{normalization}) and
the normalization is modified. 
Henceforth, we label the energies for these states as $E_n^{(-)} = \zeta_n {\cal E}_0$.

The corresponding even states must still be of the form 
$\psi_n^{(+)}(z) = Ai(z/\rho - \beta_n)$ (at least for $z>0$), 
but the appropriate boundary condition is now
that $\psi'(z=0) = Ai'(-\beta_n) = 0$, so the energy eigenvalues for these 
states are given by $E_n^{(+)} = \chi_n {\cal E}_0$, where $-\chi_n$
are the zeros of the {\it derivative} of $Ai(x)$. Using the integral
in Eqn.~(\ref{integral_diagonal_0}), we find that the appropriately normalized 
even solutions can be written in the form
\begin{equation}
\psi_n^{(+)}(z) = \frac{1}{\sqrt{2 \rho\, \chi_n} \, Ai(-\chi_n)}
Ai\left(\frac{|z|}{\rho} - \chi_n\right)
\, , 
\end{equation}
once again with an analytic result for the normalization.
A standard handbook \cite{handbook} result for the zeros of $Ai'(x)$ gives
\begin{equation}
E_n^{(+)} 
\sim  
{\cal E}_0 \left[ \frac{3\pi}{2}(n-3/4)\right]^{2/3}
= 
{\cal E}_0 \left[ \frac{3\pi}{4}((2n-1)-1/2)\right]^{2/3}
\, . 
\label{even_wkb}
\end{equation}
Comparing this to Eqn.~(\ref{quantum_bouncer_wkb}), we see that 
the energy eigenstates thus satisfy 
$E_{n}^{(+)} < E_{n}^{(-)} < E_{n+1}^{(+)}$ and interleave,
as in any Sturm-Liouville problem, and alternate in parity, as for any symmetric potential.
We note that the WKB prediction
for the symmetric linear potential is dictated by the quantization constraint
\begin{equation}
\!\!\!\!\!\!\!\!\!\!\!\!\!\!\!\!\!\!\!\!\!\!\!\!
\int_{-E_{\overline{n}}/F}^{+E_{\overline{n}}/F}
\sqrt{2m(E_{\overline{n}} -F|z|)}\,dz
 = (\overline{n}+1/4 +1/4)\hbar \pi
\qquad
\mbox{with \, $\overline{n} = 0,1,2,...$}
\end{equation}
since both turning points are now 'soft' ($C_L = C_R = 1/4$), giving
\begin{equation}
E_{\overline{n}} = \left[\frac{3\pi(\overline{n}+1/2)}{4}\right]^{3/2}
{\cal E}_0
\label{symmetric_potential_wkb}
\, . 
\end{equation}
This expression agrees with Eqn.~(\ref{quantum_bouncer_wkb}) 
(for $\overline{n} = 2n$)
and 
Eqn.~(\ref{even_wkb}) 
(for $\overline{n} = 2n-1$), as expected.

Simple calculations then show that virial theorem results in 
Eqns.~(\ref{potential_energy}) 
and (\ref{kinetic_energy}) still hold. Furthermore, 
the `diagonal' dipole matrix elements are given by
\begin{equation}
\langle \psi_n^{(+)} |z| \psi_k^{(+)} \rangle
= 0 = 
\langle \psi_n^{(-)} |z| \psi_k^{(-)} \rangle
\label{symmetric_dipole_1}
\end{equation}
because of parity constraints, while using the
integral in Eqn.~(\ref{integral_off_diagonal_1}) we find
\begin{equation}
\langle \psi_n^{(-)} |z| \psi_k^{(+)} \rangle
= -\frac{2\rho}{\sqrt{\chi_k}(\chi_k - \zeta_n)^3}
\,. 
\label{symmetric_dipole_2}
\end{equation}

Turning now to the Stark effect, the addition of an external 
constant field, $\overline{V}(z) =  \overline{F}z$, 
changes the potential asymmetrically to
\begin{equation}
\tilde{V}(z) = V(z) + \overline{V}(z) 
=
\left\{
\begin{array}{cc}
+(F+\overline{F})z & \qquad \mbox{for $z>0$} \\
-(F-\overline{F})z & \qquad \mbox{for $z<0$}
\end{array}
\right.
\, , 
\end{equation}
but the Schr\"{o}dinger equation
still supports Airy function solutions in both regions. 
The most general solution satisfying the boundary conditions 
at $\pm \infty$ is now
\begin{equation}
\psi_n(z) =
\left\{
\begin{array}{cc}
\alpha_R \, Ai(+z/\rho_R - E_n/(F_R \rho_R)) & \quad \mbox{for $z>0$} \\
\alpha_L \, Ai(-z/\rho_L - E_n/(F_L \rho_L)) & \quad \mbox{for $z<0$} 
\end{array}
\right.
\end{equation}
using the definitions
\begin{eqnarray}
F_R & = F(1+\Delta) \qquad \qquad \rho_R = & \rho(1+\Delta)^{-1/3} \\
F_L & = F(1-\Delta) \qquad \qquad \rho_L = & \rho(1-\Delta)^{-1/3}
\end{eqnarray}
where $\Delta \equiv \overline{F}/F$ and the $R/L$ notations refer to
the $z>0$ and $z<0$ regions respectively. The boundary conditions which
must now be satisfied at $z=0$ are the continuity of both $\psi_n(z)$ and
$\psi'_n(z)$ and applying these we find the eigenvalue condition
\begin{eqnarray}
\!\!\!\!\!\!\!\!\!\!\!\!\!\!\!\!\!\!\!\!\!\!
G(E_n,\Delta)
& \equiv &
 (1+\Delta)^{1/3} 
Ai\left(\frac{-E_n}{{\cal E}_0 (1-\Delta)^{2/3}}\right)
Ai'\left(\frac{-E_n}{{\cal E}_0 (1+\Delta)^{2/3}}\right)
 \nonumber \\
& & 
\qquad 
+
(1-\Delta)^{1/3} 
Ai\left(\frac{-E_n}{{\cal E}_0 (1+\Delta)^{2/3}}\right)
Ai'\left(\frac{-E_n}{{\cal E}_0 (1-\Delta)^{2/3}}\right)
= 0 
\, .
\end{eqnarray}
We note that in the absence of the perturbing  field we have
$\Delta =0$ and the eigenvalue condition reduces to
\begin{equation}
G(E_n,0) = Ai\left(\frac{-E_n}{{\cal E}_0}\right)
Ai'\left(\frac{-E_n}{{\cal E}_0}\right)
= 0
\end{equation}
which is satisfied by either $E_n/{\cal E}_0 = \chi_n$ (giving the
even states) or $\zeta_n$ (giving the odd states), as expected. 

Assuming that $\Delta << 1$, one can expand $G(E_n,\Delta)$ about 
the unperturbed values using standard Taylor series expansions of
the $Ai(x)$ and $Ai'(x)$ functions. This expansion is immensely simplified
by repeated use of the fact that the $Ai(x)$ satisfies the Airy differential
equation, namely $Ai''(x) = xAi(x)$. For example, if we need
\begin{equation}
\!\!\!\!\!\!\!\!\!\!\!\!\!\!\!\!\!\!\!\!
Ai(y+\delta) = Ai(y) + \delta Ai'(y) +\frac{\delta^2}{2!} Ai''(y)
+ \frac{\delta^3}{3!}Ai'''(y) + 
\frac{\delta^4}{4!} Ai^{(iv)}(y) + \cdots
\end{equation}
we can use the relations
\begin{eqnarray}
Ai''(y)  & = & y Ai(y) 
\label{id_2} \\
Ai'''(y) & = & [Ai''(y)]' = [yAi(y)]' = Ai(y) + yAi'(y) 
\label{id_3} \\
Ai^{(iv)}(y) & = & [Ai'''(y)]' = [Ai(y) + yAi'(y)]'
= 2Ai'(y) + yAi''(y) \nonumber \\
& = &  2Ai'(y) + y^2 Ai(y) \\
\vdots \quad & = & \qquad \qquad \vdots \nonumber 
\label{id_4}
\, \qquad. 
\end{eqnarray}
Thus, the derivatives at any order can be expressed as a combination
of $Ai(y)$ and $Ai'(y)$ (each times polynomials in $y$), 
but no higher derivatives. Then, since we are
expanding about $y = -\chi_n$ or $y= -\zeta_n$, one or the other
of these terms will vanish (either $Ai(-\zeta_n) = 0$ or $Ai'(-\chi_n) = 0$).
These simplifications leave the expansion of $G(E_n,\Delta)$
proportional to the non-vanishing term squared, multiplying a
polynomial in $\Delta$ with coefficients containing polynomials in
$\chi_n$ or $\zeta_n$. 

So, the strategy is to write 
\begin{equation}
E_n = x {\cal E}_0 (1 + R_1 \Delta + R_2\Delta^2 + R_3\Delta^3
+ \cdots)
\end{equation}
where $x = \chi_n$ or $\zeta_n$, expand $G(E_n,\Delta)$ in terms of
powers of $\Delta$, and make repeated use of the identities in 
Eqns.~(\ref{id_2}) - (\ref{id_4}). The resulting expression in powers
of $\Delta$ must vanish identically, 
so that its coefficients (which will contain the dimensionless
$R_i$ factors) 
must vanish term-by-term. 

As an example,
expanding around $E_n = \zeta_n {\cal E}_0$ for the odd states 
(for which $Ai(-\zeta_n) = 0$) gives the constraint
\begin{eqnarray}
\!\!\!\!\!\!\!\!\!\!\!\!\!
G(E_n^{(-)},\Delta) 
& = &
[Ai'(-\zeta_n)]^2
\left\{
-2\zeta_nR_1\Delta
- \frac{(14+18R_2)\zeta_n}{9}\Delta^2 \right. \nonumber \\
& & 
\qquad \qquad \qquad 
\left. 
- \frac{(3R_3 + 2R_1 - 2R_1^3\zeta_n^3)\zeta_n}{3}\Delta^3 
+\cdots\right\}
= 0 
\end{eqnarray}
up to third order in $\Delta = \overline{F}/F$. We immediately see that
$R_1 = R_3 = 0$ as dictated by the symmetry of the problem and
perturbation theory results. Recall that the first-order shift
in Eqn.~(\ref{first_order_shift}) vanishes in this case due to parity 
(since the diagonal dipole
matrix elements vanish, as in Eqn.~(\ref{symmetric_dipole_1})). The
third-order shift, given by 
\begin{equation}
\!\!\!\!\!\!\!\!\!\!\!\!\!\!\!\!\!\!\!\!\!\!\!\!\!\!\!\!\!\!\!\!\!
E_{n}^{(3)}
=
\sum_{k \neq n} \sum_{j \neq n}
\frac{
\langle n |\overline{V}(z)| k \rangle
\langle k |\overline{V}(z)| j \rangle
\langle j |\overline{V}(z)| n \rangle}
{
(E_n^{(0)} - E_{k}^{(0)})
(E_n^{(0)} - E_{j}^{(0)})
}
 -
\langle n|\overline{V}(z)|n \rangle
\sum_{k \neq n}
\frac{|\langle n |\overline{V}(z)|k\rangle|^2}{(E_{n}^{(0)} - E_{k}^{(0)})^2}
\, , 
\end{equation}
can then be seen to explicitly vanish as well for similar reasons.

The new non-trivial result is that $R_2 = -7/9$, 
so that the Stark shift for the odd states is given by
\begin{equation}
E_n^{(-,2)} = - \frac{7}{9} \left(\frac{\overline{F}}{F}\right)^2 
(\zeta_n {\cal E}_0)
= 
- \frac{7}{9} \left(\frac{\overline{F}}{F}\right)^2 \, E_n^{(-,0)}
\label{big_new_odd}
\,. 
\end{equation}
In exactly the same manner, expanding about $E_n = \chi_n {\cal E}_0$, we
find the corresponding result for the Stark shift for the even states, namely
\begin{equation}
E_n^{(+,2)} = - \frac{5}{9} \left(\frac{\overline{F}}{F}\right)^2 
(\chi_n {\cal E}_0)
=
- \frac{5}{9} \left(\frac{\overline{F}}{F}\right)^2 \, E_n^{(+,0)}
\label{big_new_even}
\,. 
\end{equation}

If we include the perturbing potential, the WKB prediction for the energy 
eigenvalues is now determined by the quantization condition
\begin{equation}
\sqrt{2m} \int_{a_{-}}^{a_{+}}
\sqrt{\tilde{E}_n - (F|z| + \overline{F}z)}
\, dz = (n+1/2)\pi \hbar
\end{equation}
where $n = 0,1,2,...$ and the classical turning points are
\begin{equation}
a_{\pm} = \pm \frac{E_n}{(F \pm \overline{F})}
\,. 
\end{equation}
The resulting approximation for the energy eigenvalues is found to be 
\begin{equation}
\tilde{E}_n = E_n\left(1- \frac{\overline{F}^2}{F^2}\right)^{2/3}
\approx
E_n\left( 1 - \frac{2}{3} \Delta^2+ \cdots \right)
\end{equation}
where the $E_n$ are given by the $\overline{F}=0$ WKB result in
Eqn.~(\ref{symmetric_potential_wkb}).
In this approximate approach, the first- and third-order terms (in $\Delta$)
do indeed vanish, while the second-order Stark shift is given by
\begin{equation}
E_n^{(2)} = -\frac{2}{3}E_n \Delta^2 
= -\frac{6}{9} \left(\frac{\overline{F}}{F}\right)^2\, E_n^{(0)}
\end{equation}
with no dependence on parity, 
since that concept plays no role in the semi-classical WKB approach.
(The Stark shifts in various power-law
potentials have been examined in Ref.~\cite{polar} using just such a WKB
approach, where this last result was first noted.)

Interestingly then, we find that the exact second-order results 
in Eqns.~(\ref{big_new_odd}) 
and (\ref{big_new_even}) bracket the approximate WKB result 
and in fact give it as the 'average' effect, but they do depend on the
parity of the unperturbed eigenstate. Both approaches give the correct
dependence as being proportional to the unperturbed energy eigenvalue.

Finally, in the spirit of Ref.~\cite{belloni_robinett_airy_sum_rules},
we can use the standard second-order perturbation theory expression 
in Eqn.~(\ref{general_second_order_shift}) and the dipole matrix elements in 
Eqns.~(\ref{symmetric_dipole_1}) and (\ref{symmetric_dipole_2}) to find
\begin{eqnarray}
E_n^{(-,2)} & = & 
- 4\left(\frac{\overline{F}}{F}\right)^2 {\cal E}_0 
\left[\sum_{k} \frac{1}{\chi_k(\chi_k - \zeta_n)^7}\right] 
\\
E_n^{(+,2)} & = & 
- 4\left(\frac{\overline{F}}{F}\right)^2 {\cal E}_0 
\left[\sum_{k} \frac{1}{\chi_n(\zeta_k - \chi_n)^7}\right] 
\end{eqnarray}
which leads to two new constraints on the $\chi_n,\zeta_n$, namely
\begin{equation}
\!\!\!\!\!\!\!\!\!\!\!\!
\left(\frac{1}{\chi_n}\right)\sum_{k} \frac{1}{(\zeta_k - \chi_n)^7} = +\frac{5 \chi_n}{36}
\qquad
\mbox{and}
\qquad
\sum_{k} \frac{1}{\chi_k(\chi_k - \zeta_n)^7} = +\frac{7 \zeta_n}{36}
\end{equation}
both of which are easily confirmed numerically using programs such as
\mathmath. (In fact, in the spirit of Ref.~\cite{bouncing_ball}, we
actually found numerical evidence for these expressions first, 
and then derived them in closed form using the approach followed above.)

We note that
the expansion of $G(E_n,\Delta)$ can be continued to essentially arbitrarily
high order, by extending the Airy function identities in
Eqns.~(\ref{id_2}) - (\ref{id_4}). 
Comparing the next non-trivial result ($E_n^{(\pm,4)}$) to standard
expressions for the 4th-order energy shift (see, e.g., Ref.~\cite{4th_order} 
for the form), for example, yields a complicated constraint on the $\zeta_n,\chi_n$
in the form of a triple infinite summation.

\section{Discussion and conclusions}
\label{sec:conclusions}

We have derived closed form expressions for the second-order energy
shift due to a constant external field (Stark effect) for two model
systems, using simple properties of the solutions of the Schr\"{o}dinger 
equation for the linear potential relevant for both problems.
 These expressions add to the handful (almost literally) of
exact results for the Stark shifts which are available 
for mathematically tractable model
systems in quantum mechanics (including the hydrogen atom, 
harmonic oscillator, and infinite well.)

The results in Eqn.~(\ref{big_new_odd}) and (\ref{big_new_even}) 
for the symmetric linear potential are very similar, but not identical to, 
a simple WKB prediction, 
but with an explicit dependence on the parity of the state.
In contrast, the exact results for the quantum
bouncer are trivially implemented and agree with both the WKB approach
as well as perturbation theory (up to at least 3rd order 
\cite{belloni_robinett_airy_sum_rules}.)

We can now compare the results for the symmetric linear potential
to two other familiar cases, such as the harmonic oscillator result in 
Eqn.~(\ref{harmonic_oscillator_stark}) or the
infinite well in Eqn.~(\ref{infinite_well_stark}) where the
first order results both vanish as well.
The form of the second-order shift in Eqn.~(\ref{general_second_order_shift})
requires that the result for $E_n^{(2)}$ for the ground state of any system
(assumed non-degenerate) is always negative and that is indeed confirmed. 
Similarly to the harmonic oscillator, 
the second-order shifts for all higher states here are also  negative, 
but are proportional to the initial energies, and not constant. This
result is conceptually consistent with the WKB discussions in 
Ref.~\cite{polar} and has the dependence on quantum number predicted there, 
but the simple WKB approach does not correctly distinguish between the 
results for different parity quantum solutions.

One well-known textbook author \cite{saxon} in discussing second-order
perturbation theory has noted that ``{\it ...the response of any system to
a perturbing force is to deform at the expense of that force 
and hence to \underline{decrease} 
the potential energy of interaction}''. He concludes, however, that 
``{\it ...on physical grounds we \underline{expect} the second-order 
correction to lower the energy in general but have \underline{proved} 
this to be the case only for the ground state.}'' Our analysis provides
other examples of where this intuition is proved correct, with the 
explicit form derived in closed form for all quantum states in the system.

Motivated by such observations, 
the discussion in Ref.~\cite{polar} used WKB approximations
to examine the 'cross-over' behavior of power-law potentials of the
form in Eqn.~(\ref{power_law}) where for small $k$ ($k=1,2$ for the
symmetric linear potential and oscillator) the second-order shifts were
all negative, while for large $k$ ($k\rightarrow \infty$ for the
infinite well) all values of $E_n^{(2)}$ were positive, except as required
for the ground state. One wanted to see just how `stiff' the potential 
had to be for the large $n$ (or even $n>1$) behavior of the 
second-order shifts to change qualitatively in sign.
The  results derived here provide additional evidence of the 
validity of such a WKB approach to that general question, but also give 
more examples of exact closed-form expressions 
for the Stark effect, and how they depend on the
details of the quantum solutions, in this case their parity.

In our review of the properties of the quantum bouncer and symmetric linear
potential, we have collected several existing closed-form results related
to the two systems, while providing a number of new analytic examples.
Each  of these results depend on relatively straightforward identities (such as
the integrals in the Appendix) or other properties of the Airy functions,
all directly related to the simple form of the differential equation they 
solve.  Many of the `standard' model quantum mechanical systems
which are of relevance to important experiments (Coulomb potential, harmonic
oscillator, spherical harmonics) require moderately sophisticated
mathematical physics techniques to obtain even some of the simplest results;
one exception is the infinite well, which relies only on 
simple integrals involving familiar trigonometric functions. Another
often discussed problem is the Stark effect for the plane rotator
\cite{rot_1} - \cite{rot_4} for which the application of perturbation theory 
is less than straightforward, so much so that a number of textbooks have
evidently gotten it wrong \cite{rot_5} - \cite{rot_9}. The correct results
in that case can also be obtained by explicit expansion of 
the exact eigenvalue condition, but this requires knowing the 
properties of Mathieu function.  In contrast to
these cases, the results shown here are all derivable
(or can be easily checked by simple differentiation) 
using only the fact that $Ai''(x) = xAi(x)$ and
little else, making them very accessible to students of quantum 
mechanics at the undergraduate level, without the need for advanced 
mathematical background. The fact that they satisfy the Schr\"{o}dinger 
equation is all that's needed.

Our focus on these analytic solutions is motivated by recent discussions 
of new mathematical results \cite{bouncing_ball} - \cite{goodmanson}
involving Airy functions using just such simple identities.
Many of these results have been in the research 
literature for decades \cite{gordon}, \cite{albright}, but are only 
now finding use because of the application of this once 'abstract'
problem to  novel physical systems \cite{neutron_bound_states} - 
\cite{bouncing_photon}. 

As recently as 1992, one author \cite{bouncer_5}
was led to describe the analytic solutions of the quantum bouncer 
in the following way:
``{\it This technique, which leads to Airy functions, tends to be
algebraically tedious and not particularly instructive; it is best
reserved for more important potentials, such as the harmonic oscillator
or Coulomb potentials.}'' Given the increased physical relevance of this
model system,  and its mathematically tractable nature, 
we argue both for its utility as a useful pedagogical example
and for an enhanced position in the pantheon 
of canonical quantum mechanical showcase problems.

\section{Appendix}

We collect here some of the useful identities involving integrals over
Airy functions needed in this paper. Versions of these results have been
derived in Refs.~\cite{gordon} and \cite{albright}.  In each case,
the identities are most easily proved by direct differentiation of the
right hand sides. (For example, Albright \cite{albright} notes 
that ``{\it Like most results in integral calculus, they are easier 
to check than to obtain.}'')

We first assume that $A(x)$ and $B(x)$ are any two 
linearly-independent solutions of the Airy differential equation,
\begin{equation}
\!\!\!\!\!\!\!\!\!\!\!\!\!\!\!\!\!\!\!\!\!\!\!\!
\!\!\!\!\!\!\!\!\!\!\!\!\!
A(x-\beta)  =  a Ai(x-\beta) + b Bi(x-\beta) 
\quad
\mbox{and}
\quad
B(x-\beta)  =  c Ai(x-\beta) + d Bi(x-\beta) 
\end{equation}
where
\begin{equation}
\!\!\!\!\!\!\!\!\!\!\!\!
A''(x) = (x-\beta) A(x-\beta)
\qquad
\mbox{and}
\qquad
B''(x-\beta) = (x-\beta) B(x-\beta)
\end{equation}
where we consider the shifted arguments needed for solutions of
the quantum bouncer and symmetric linear potential. We then 
have the following identities involving indefinite integrals of
products of the $A,B$, their derivatives, and various moments.

The first identity below is needed for the normalization of wavefunctions, 
while the second and third are used to evaluate expectation values.
\begin{eqnarray}
\!\!\!\!\!\!\!\!\!\!\!\!\!\!\!\!\!\!\!\!\!\!\!\!\!\!
\int \, A(x-\beta) \, B(x-\beta)\, dx
& = & 
(x-\beta) A(x-\beta) B(x-\beta) - A'(x-\beta) B'(x-\beta) 
\label{integral_diagonal_0} 
\end{eqnarray}

\begin{eqnarray}
\!\!\!\!\!\!\!\!\!\!\!\!\!\!\!\!\!\!\!\!\!\!\!\!\!\!\!\!\!\!\!\!
\int \, x\, A(x-\beta) \, B(x-\beta)\, dx
& = &
\frac{1}{3} (x^2 +  \beta x - 2\beta^2) A(x-\beta) B(x-\beta) \nonumber \\
& &
\qquad
+\frac{1}{6}\left\{A'(x-\beta) B(x-\beta) + A(x-\beta) B'(x-\beta)\right\}
\nonumber \\
& & 
\qquad \qquad
- \frac{1}{3} (x+2\beta)A'(x-\beta) B'(x-\beta) 
\label{integral_diagonal_1} 
\end{eqnarray}

\begin{eqnarray}
\!\!\!\!\!\!\!\!\!\!\!\!\!\!\!\!\!\!\!\!\!\!\!\!\!\!\!\!\!\!\!\!\!
\int \, x^2\, A(x-\beta) \, B(x-\beta)\, dx
& = &
\frac{1}{15} (3x^3 + \beta x^2 + 4\beta^2x - 8\beta^3 -3)
A(x-\beta) B(x-\beta) \nonumber \\
&& 
+ \frac{1}{15}(3x + 2\beta)
\left\{A'(x-\beta) B(x-\beta) + A(x-\beta) B'(x-\beta)\right\}
\nonumber \\
&&
\qquad 
- \frac{1}{15}(3x^2 + 4\beta x + 8\beta^2) A'(x-\beta) B'(x-\beta)
\label{integral_diagonal_2} 
\, . 
\end{eqnarray}

The following identity involving derivatives is useful for evaluation of the
expectation value of the kinetic energy operator and can be derived from
one in Ref.~\cite{albright} by a simple change of variables.
\begin{eqnarray}
\!\!\!\!\!\!\!\!\!\!\!\!\!\!\!\!\!\!\!\!\!\!\!
\int A'(x-\beta) \, B'(x-\beta)\, dx
& = & \frac{1}{3}
\left[
\left\{A'(x-\beta) B(x-\beta) + A(x-\beta) B'(x-\beta)\right\} 
\right. \nonumber \\
& & 
\!\!\!\!\!\!\!\!\!\!\!\!\!\!\!\!\!\!\!\!\!\!\!\!\!\!\!\!\!\!\!\!\!\!\!\!\!\!\!\!\!\!\!\!
\left. + (x-\beta) A'(x-\beta) B'(x-\beta) - (x-\beta)^2 A(x-\beta) B(x-\beta)
\right]
\label{integral_momentum} 
\, . 
\end{eqnarray}

We often need to consider integrals involving solutions with two
different 'shifts' ($\beta_1 \neq \beta_2$), 
for example in the confirmation of the orthogonality
of eigenstates (first identity) or the evaluation of dipole matrix 
elements, as in Eqns.~(\ref{bouncer_dipole_matrix_elements}) 
and (\ref{symmetric_dipole_2}).
\begin{eqnarray}
\!\!\!\!\!\!\!\!\!\!\!\!\!\!\!\!\!\!\!\!\!\!\!\!\!\!\!\!\!
\int \, A(x-\beta_1) \, B(x-\beta_2)\, dx
& = & \frac{1}{(\beta_2-\beta_1)}
\left\{
A'(x-\beta_1) B(x-\beta_2)\right. \nonumber \\
& &
\left.
\qquad \qquad \qquad 
- A(x-\beta_1) B'(x-\beta_2) \right\} 
\label{integral_off_diagonal_0}
\end{eqnarray}

\begin{eqnarray}
\!\!\!\!\!\!\!\!\!\!\!\!\!\!\!\!\!\!\!\!\!\!\!\!\!\!\!\!\!\!\!\!\!\!\!\!\!\!\!\!\!\!\!\!\!
\int \, x\, A(x-\beta_1) \, B(x-\beta_2)\, dx
& = & 
\frac{(\beta_1 + \beta_2 - 2x)}{(\beta_1 - \beta_2)^2}
A(x-\beta_1) B(x-\beta_2) \nonumber \\
& & 
\!\!\!\!\!\!\!\!\!\!\!\!\!\!\!\!\!\!\!\!\!\!\!\!\!\!\!\!\!\!\!\!\!\!\!\!\!\!
+
\left\{ 
\frac{x}{(\beta_2 - \beta_1)}
+ 
\frac{2}{(\beta_2 - \beta_1)^3}
\right\}
\left\{
A'(x-\beta_1) B(x-\beta_2) - A(x-\beta_1) B'(x-\beta_2)
\right\} \nonumber \\
& & 
\qquad
\frac{2}{(\beta_1-\beta_2)^2}
A'(x-\beta_1) B'(x-\beta_2)
\label{integral_off_diagonal_1} 
\, . 
\end{eqnarray}

\end{document}